# FIRST INTEGRALS FOR THE VOLTERRA CHAIN

## Belskiy D. V.

*Institute of Mathematics of the National Academy of Sciences of Ukraine*
*Tereschenkivska St., 3, Kyiv, 01024, Ukraine*
*e-mail: oiop120@gmail.com*

New determinant equalities were obtained based on the Wronskian formulas for a particular solution of the Volterra chain. Using the relationship between the Toda and Volterra chains, new first integrals for the Volterra chain are calculated using the first integrals for the Toda chain. Using the first integrals, a periodic Volterra chain with a period of five was considered.

**1. Introduction.** This article considers the Volterra chain
$$u'_i = u_i(u_{i+1} - u_{i-1}), \ i \in \mathbb{Z}, \tag{1}$$
which, like the Toda chain [1, 2], is a difference analogue of the Korteweg–de Vries equation [3, 4]. In the case of $u_i > 0$, this system, in particular, arises when studying the fine structure of the spectra of Langmuir oscillations in plasma, which is why it is sometimes called the Langmuir chain [3]. The Volterra chain has been studied in many works (see [5-8] and the literature cited there).

In the second part of this note, new algebraic determinant identities are obtained from one particular solution of the Volterra chain on the half-axis $i \geq 0$ using the Wronskian formulas in [9]. In the third part, using the connection between the solutions of the Volterra and Toda chains in [8], new first integrals for the Volterra chain are calculated based on the first integrals for the Toda chain in [10]; the latter complements the corresponding results in [2, 9].

The periodic Volterra chain was studied in general in [5, 6]. For periods 3 and 4, it was also investigated in [7, 11] using first integrals, which made it possible to find and solve differential equations for the desired functions. In the fourth part of this note, the Volterra chain with period 5 is also studied using first integrals, which allows us to reduce the dimensionality of the system and, in some cases, to calculate and solve differential equations for the desired functions.

**2. Determinant identities.** Let us prove the following lemma.

**Lemma.** *The following determinant identities*

$$\begin{vmatrix} 1 & \dfrac{(-c_1)_1}{(-a)^1} & \dfrac{(-c_1)_2}{(-a)^2} & \cdots & \dfrac{(-c_1)_k}{(-a)^k} \\ \dfrac{(-c_1)_1}{(-a)^1} & \dfrac{(-c_1)_2}{(-a)^2} & & & \vdots \\ \dfrac{(-c_1)_2}{(-a)^2} & & \ddots & & \dfrac{(-c_1)_{2k-2}}{(-a)^{2k-2}} \\ \vdots & & & \dfrac{(-c_1)_{2k-2}}{(-a)^{2k-2}} & \dfrac{(-c_1)_{2k-1}}{(-a)^{2k-1}} \\ \dfrac{(-c_1)_k}{(-a)^k} & \cdots & \dfrac{(-c_1)_{2k-2}}{(-a)^{2k-2}} & \dfrac{(-c_1)_{2k-1}}{(-a)^{2k-1}} & \dfrac{(-c_1)_{2k}}{(-a)^{2k}} \end{vmatrix} =$$

$$= a^{-k(k+1)} 1!2!...k! \prod_{j=1}^{k} (j-1-c_1)^{k+1-j},$$

$$\begin{vmatrix} \dfrac{(-c_1)_1}{(-a)^1} & \dfrac{(-c_1)_2}{(-a)^2} & \dfrac{(-c_1)_3}{(-a)^3} & \cdots & \dfrac{(-c_1)_{k+1}}{(-a)^{k+1}} \\ \dfrac{(-c_1)_2}{(-a)^2} & \dfrac{(-c_1)_3}{(-a)^3} & & & \vdots \\ \dfrac{(-c_1)_3}{(-a)^3} & & \ddots & & \dfrac{(-c_1)_{2k-1}}{(-a)^{2k-1}} \\ \vdots & & & \dfrac{(-c_1)_{2k-1}}{(-a)^{2k-1}} & \dfrac{(-c_1)_{2k}}{(-a)^{2k}} \\ \dfrac{(-c_1)_{k+1}}{(-a)^{k+1}} & \cdots & \dfrac{(-c_1)_{2k-1}}{(-a)^{2k-1}} & \dfrac{(-c_1)_{2k}}{(-a)^{2k}} & \dfrac{(-c_1)_{2k+1}}{(-a)^{2k+1}} \end{vmatrix} =$$

$$= c_1^{k+1} \dfrac{1}{a^{(k+1)^2}} 1!2!\ldots k! \prod_{j=1}^{k} (j - c_1)^{k+1-j}, \quad k \geq 1 \qquad (2)$$

hold, where $(\alpha)_n = \alpha(\alpha+1)\ldots(\alpha+n-1)$, $n \in \mathbb{N}$, $(\alpha)_0 = 1$, is the Pochhammer symbol; $a \neq 0$ and $c_1$ are some real numbers.

**Proof.** Let us consider a particular solution of the Volterra chain on a semi-axis $i \geq 0$:

$$u_{2k} = -\dfrac{k}{x+a}, \quad u_{2k+1} = \dfrac{c_1 - k}{x+a}, \quad k \geq 0,$$

where $a \neq 0$ and $c_1$ are some real numbers. Let's write the function $u_1$ as the following fraction

$$\dfrac{c_1}{x+a} = u_1 = \dfrac{f'}{f},$$

$$f(x) = \left(1 + \dfrac{x}{a}\right)^{c_1} = 1 + \sum_{n=1}^{+\infty} \dfrac{(-c_1)_n}{(-a)^n} \dfrac{x^n}{n!},$$

where the power series is the binomial series (see, for example, [12], p. 285). Then, according to statement 7 in [9], the determinant identities (2) are satisfied.

The lemma is proved.

In the case $a = c_1 = -1$, the first equality in (2) is the first point of fact 3.16.28 in [13], on page 334. We also note that the Wronskian formulas in [9] for solutions of the Volterra chain on the semiaxis $i \geq 0$, with the help of which the lemma was proved, are of little use for practical calculation of the solution, but *they do allow one to obtain new algebraic determinant identities*.

**3. First integrals for the Volterra chain.** In this section, we will need first integrals for the Toda chain to calculate new first integrals for the Volterra chain using the relationship between the Toda and Volterra chains. Therefore, we will write the Toda chain in the form from [10]:

$$x_i' = U_i, \quad U_i' = C\left[e^{-(x_i - x_{i-1})} - e^{-(x_{i+1} - x_i)}\right], \qquad (3)$$

where $C$ is a constant, $i \in \mathbb{Z}$. We define the functions

$$X_i = Ce^{-(x_{i+1} - x_i)},$$

then system (3) takes the form

$$X_i' = (U_i - U_{i+1}) X_i, \quad U_i' = X_{i-1} - X_i. \qquad (4)$$

Let us make a small remark about the first integrals of the Toda chain. *In [10] a connection is indicated between the first integrals of two types $I_m$ and $J_m$ of system (4) through Newton's formulas for symmetric polynomials* (see, for example, (1.4-11) or (1.6-5) in [14]), if in these formulas the polynomials $S_m$ and $s_m$ are replaced by the first integrals $I_m$ and $mJ_m$ from [10], respectively. But simple calculations, when the period of system (4) is $n = 2$, show that Newton's formulas (1.4-11) in [14] for $k > n$ do not ensure the indicated connection.

From the solution of the Volterra chain, one can obtain two solutions of the Toda chain (see, for example, [8], systems (3.3.28), (3.3.29)) using the following equalities

$$X_i = u_{2i+2} u_{2i+1}, \ U_i = -(u_{2i+1} + u_{2i}); \tag{5}$$

$$\hat{X}_i = u_{2i+1} u_{2i}, \ \hat{U}_i = -(u_{2i} + u_{2i-1}). \tag{6}$$

However, most solutions of the Toda chain cannot be expressed in terms of solutions of the Volterra chain using formulas (5), (6). If a solution of the Volterra chain has a period $2n$, then the solutions of the Toda chain (5), (6) have a period of $n$. In the case of an odd period $n = 2l+1$, $l \geq 0$, of a solution of the Volterra chain, the solutions of the Toda chain (5), (6) are related by the relations

$$X_{i+l} = \hat{X}_i, \ U_{i+l} = \hat{U}_i.$$

The first integrals for the Volterra chain (1) with period $n$

$$H_1 = \sum_{i=1}^{n} u_i, \ H_2 = \sum_{i=1}^{n} \left( \frac{1}{2} u_i^2 + u_i u_{i+1} \right)$$

were obtained in [9]. We calculate the following first integral for the Volterra chain. To do this, we make the substitution of variables (5) in the formula

$$J_3 = \sum_{i=1}^{n} \left[ \frac{1}{3} U_i^3 + (U_i + U_{i+1}) X_i \right]$$

(see [10]), where the solution to the Volterra chain has period $n$. This results in the equality

$$-J_3 = 2 \sum_{i=1}^{n} \left( \frac{1}{3} u_i^3 + u_i^2 u_{i+1} + u_i u_{i+1}^2 + u_i u_{i+1} u_{i+2} \right) \overset{df}{=} 2 H_3.$$

Let us differentiate the $i$-th term of the last sum by virtue of the Volterra system (1):

$$\frac{d}{dt} \left( \frac{1}{3} u_i^3 + u_i^2 u_{i+1} + u_i u_{i+1}^2 + u_i u_{i+1} u_{i+2} \right) =$$

$$= u_i u_{i+1}^3 + 2 u_i u_{i+1}^2 u_{i+2} + u_i u_{i+1} u_{i+2}^2 + u_i u_{i+1} u_{i+2} u_{i+3} -$$

$$- \left( u_{i-1} u_i^3 + 2 u_{i-1} u_i^2 u_{i+1} + u_{i-1} u_i u_{i+1}^2 + u_{i-1} u_i u_{i+1} u_{i+2} \right).$$

From the last equality and the condition $u_{i+n} = u_i$, $i \in \mathbb{Z}$, we obtain the identity $H_3' \equiv 0$.

Similarly, we calculate the first integral of the Volterra chain using

$$J_4 = \sum_{i=1}^{n} \left[ \frac{1}{4} U_i^4 + \left( U_i^2 + U_i U_{i+1} + U_{i+1}^2 \right) X_i + \frac{1}{2} X_i^2 + X_i X_{i+1} \right]$$

from [10]. To do this, we again make the substitution of variables (5) in the last formula, where the Volterra chain solution has a period of $n$, and obtain the equality

$$J_4 = 2 \sum_{i=1}^{n} \left( \frac{1}{4} u_i^4 + u_i^3 u_{i+1} + \frac{3}{2} u_i^2 u_{i+1}^2 + u_i^2 u_{i+1} u_{i+2} + u_i u_{i+1}^3 + 2 u_i u_{i+1}^2 u_{i+2} + \right.$$

$$\left. + u_i u_{i+1} u_{i+2}^2 + u_i u_{i+1} u_{i+2} u_{i+3} \right) \overset{df}{=} 2 H_4.$$

Let us again differentiate the $i$-th term of the last sum using the Volterra system:

$$\frac{d}{dt} \left( \frac{1}{4} u_i^4 + u_i^3 u_{i+1} + \frac{3}{2} u_i^2 u_{i+1}^2 + u_i^2 u_{i+1} u_{i+2} + u_i u_{i+1}^3 + 2 u_i u_{i+1}^2 u_{i+2} + \right.$$

$$\left. + u_i u_{i+1} u_{i+2}^2 + u_i u_{i+1} u_{i+2} u_{i+3} \right) =$$

$$= u_i u_{i+1}^4 + 3 u_i u_{i+1}^3 u_{i+2} + 3 u_i u_{i+1}^2 u_{i+2}^2 + 2 u_i u_{i+1}^2 u_{i+2} u_{i+3} + u_i u_{i+1} u_{i+2}^3 + 2 u_i u_{i+1} u_{i+2}^2 u_{i+3} +$$

$$+ u_i u_{i+1} u_{i+2} u_{i+3}^2 + u_i u_{i+1} u_{i+2} u_{i+3} u_{i+4} -$$

$$- \left( u_{i-1} u_i^4 + 3 u_{i-1} u_i^3 u_{i+1} + 3 u_{i-1} u_i^2 u_{i+1}^2 + 2 u_{i-1} u_i^2 u_{i+1} u_{i+2} + u_{i-1} u_i u_{i+1}^3 + 2 u_{i-1} u_i u_{i+1}^2 u_{i+2} + \right.$$

$$\left. + u_{i-1} u_i u_{i+1} u_{i+2}^2 + u_{i-1} u_i u_{i+1} u_{i+2} u_{i+3} \right).$$

As in the previous case, from the last equality and the condition $u_{i+n} = u_i$, $i \in \mathbb{Z}$, we obtain the identity $H_4' \equiv 0$.

Finally, we calculate one more first integral of the Volterra chain using

$$J_5 = \sum_{i=1}^{n} \left[ \frac{1}{5} U_i^5 + \left( U_i^3 + U_i^2 U_{i+1} + U_i U_{i+1}^2 + U_{i+1}^3 \right) X_i + \left( U_i + U_{i+1} \right) X_i^2 + \right.$$

$$+\left(U_i + 2U_{i+1} + U_{i+2}\right)X_i X_{i+1}\Big]$$

from [10]. To do this, we make the substitution of variables (5) in the last formula, where the Volterra chain solution has a period of $n$, and obtain the equality

$$-J_5 = 2\sum_{i=1}^{n}\left(\frac{1}{5}u_i^5 + u_i^4 u_{i+1} + 2u_i^3 u_{i+1}^2 + u_i^3 u_{i+1}u_{i+2} + 2u_i^2 u_{i+1}^3 + 3u_i^2 u_{i+1}^2 u_{i+2} + \right.$$
$$+ u_i^2 u_{i+1}u_{i+2}^2 + u_i^2 u_{i+1}u_{i+2}u_{i+3} + u_i u_{i+1}^4 + 3u_i u_{i+1}^3 u_{i+2} + 3u_i u_{i+1}^2 u_{i+2}^2 +$$
$$+ 2u_i u_{i+1}^2 u_{i+2}u_{i+3} + u_i u_{i+1}u_{i+2}^3 + 2u_i u_{i+1}u_{i+2}^2 u_{i+3} + u_i u_{i+1}u_{i+2}u_{i+3}^2 +$$
$$\left. + u_i u_{i+1}u_{i+2}u_{i+3}u_{i+4}\right)\stackrel{df}{=} 2H_5.$$

Let us differentiate the $i$-th term of the last sum using the Volterra system:

$$\frac{d}{dt}\left(\frac{1}{5}u_i^5 + u_i^4 u_{i+1} + 2u_i^3 u_{i+1}^2 + u_i^3 u_{i+1}u_{i+2} + 2u_i^2 u_{i+1}^3 + 3u_i^2 u_{i+1}^2 u_{i+2} + u_i^2 u_{i+1}u_{i+2}^2 + \right.$$
$$+ u_i^2 u_{i+1}u_{i+2}u_{i+3} + u_i u_{i+1}^4 + 3u_i u_{i+1}^3 u_{i+2} + 3u_i u_{i+1}^2 u_{i+2}^2 + 2u_i u_{i+1}^2 u_{i+2}u_{i+3} +$$
$$\left. + u_i u_{i+1}u_{i+2}^3 + 2u_i u_{i+1}u_{i+2}^2 u_{i+3} + u_i u_{i+1}u_{i+2}u_{i+3}^2 + u_i u_{i+1}u_{i+2}u_{i+3}u_{i+4}\right) =$$
$$= u_i u_{i+1}^5 + 4u_i u_{i+1}^4 u_{i+2} + 6u_i u_{i+1}^3 u_{i+2}^2 + 3u_i u_{i+1}^3 u_{i+2}u_{i+3} + 4u_i u_{i+1}^2 u_{i+2}^3 +$$
$$+ 6u_i u_{i+1}^2 u_{i+2}^2 u_{i+3} + 2u_i u_{i+1}^2 u_{i+2}u_{i+3}^2 + 2u_i u_{i+1}^2 u_{i+2}u_{i+3}u_{i+4} + u_i u_{i+1}u_{i+2}^4 +$$
$$+ 3u_i u_{i+1}u_{i+2}^3 u_{i+3} + 3u_i u_{i+1}u_{i+2}^2 u_{i+3}^2 + 2u_i u_{i+1}u_{i+2}^2 u_{i+3}u_{i+4} + u_i u_{i+1}u_{i+2}u_{i+3}^3 +$$
$$+ 2u_i u_{i+1}u_{i+2}u_{i+3}^2 u_{i+4} + u_i u_{i+1}u_{i+2}u_{i+3}u_{i+4}^2 + u_i u_{i+1}u_{i+2}u_{i+3}u_{i+4}u_{i+5} -$$
$$- \left(u_{i-1}u_i^5 + 4u_{i-1}u_i^4 u_{i+1} + 6u_{i-1}u_i^3 u_{i+1}^2 + 3u_{i-1}u_i^3 u_{i+1}u_{i+2} + 4u_{i-1}u_i^2 u_{i+1}^3 +\right.$$
$$+ 6u_{i-1}u_i^2 u_{i+1}^2 u_{i+2} + 2u_{i-1}u_i^2 u_{i+1}u_{i+2}^2 + 2u_{i-1}u_i^2 u_{i+1}u_{i+2}u_{i+3} + u_{i-1}u_i u_{i+1}^4 +$$
$$+ 3u_{i-1}u_i u_{i+1}^3 u_{i+2} + 3u_{i-1}u_i u_{i+1}^2 u_{i+2}^2 + 2u_{i-1}u_i u_{i+1}^2 u_{i+2}u_{i+3} + u_{i-1}u_i u_{i+1}u_{i+2}^3 +$$
$$\left. + 2u_{i-1}u_i u_{i+1}u_{i+2}^2 u_{i+3} + u_{i-1}u_i u_{i+1}u_{i+2}u_{i+3}^2 + u_{i-1}u_i u_{i+1}u_{i+2}u_{i+3}u_{i+4}\right).$$

Again, from the last equality and the condition $u_{i+n} = u_i$, $i \in \mathbb{Z}$, we obtain the identity $H_5' \equiv 0$.

Substituting the change of variables (6) into $J_m$ leads to the same results. Note that the obtained first integrals $H_m$ are not independent in the general case, for example, for periods $n = 4, 5$ the equality

$$\begin{vmatrix} H_1 & 1 & 0 \\ 2H_2 & H_1 & 2 \\ 3H_3 & 2H_2 & H_1 \end{vmatrix} = 0$$

holds.

In [9] the following first integral was also obtained for a Volterra chain with period $n$:

$$H_0 = \sum_{i=1}^{n}\ln u_i.$$

For an even period $n = 2l$, $l \geq 1$, $H_0$ splits into two first integrals

$$H_{0,1} = \sum_{i=1}^{l}\ln u_{2i}, \quad H_{0,2} = \sum_{i=0}^{l-1}\ln u_{2i+1}.$$

The construction of first integrals for infinite Volterra chains with $i \in \mathbb{Z}$ and $i \geq 0$ can be found in [9].

**4. Volterra chain with period 5.** We write the Volterra chain with period $n = 5$ in expanded form

$$\begin{cases} u_1' = u_1(u_2 - u_5) \\ u_2' = u_2(u_3 - u_1) \\ u_3' = u_3(u_4 - u_2). \\ u_4' = u_4(u_5 - u_3) \\ u_5' = u_5(u_1 - u_4) \end{cases} \qquad (7)$$

And we write out the first three integrals of system (7):

$$u_1 u_2 u_3 u_4 u_5 = B, \qquad (8)$$

$$u_1 + u_2 + u_3 + u_4 + u_5 = C, \qquad (9)$$

$$\frac{1}{2}\left(H_1^2 - 2H_2\right) = u_1 u_3 + u_1 u_4 + u_2 u_4 + u_2 u_5 + u_3 u_5 = D, \qquad (10)$$

where $B$, $C$, $D$ are some constants. From (8) and (9) we obtain the following relationships between the sought functions

$$u_2 = \frac{1}{2}\left[-(u_1 + u_3 + u_4 - C) \pm \sqrt{(u_1 + u_3 + u_4 - C)^2 - \frac{4B}{u_1 u_3 u_4}}\right], \qquad (11)$$

$$u_5 = \frac{1}{2}\left[-(u_1 + u_3 + u_4 - C) \mp \sqrt{(u_1 + u_3 + u_4 - C)^2 - \frac{4B}{u_1 u_3 u_4}}\right]. \qquad (12)$$

The choice of the required sign in the last two equalities is determined by the initial values of the sought functions. Substituting expressions (11) and (12) into identity (10), we obtain

$$D = \frac{1}{2}(u_3 + u_4)\left[u_1 - (u_3 + u_4) + C\right] \mp$$

$$\mp \frac{1}{2}(u_3 - u_4)\sqrt{(u_1 + u_3 + u_4 - C)^2 - \frac{4B}{u_1 u_3 u_4}} + \frac{B}{u_1 u_3 u_4},$$

$$\left(u_1\left\{D - \frac{1}{2}(u_3 + u_4)\left[u_1 - (u_3 + u_4) + C\right]\right\} - \frac{B}{u_3 u_4}\right)^2 =$$

$$= \frac{1}{4}(u_3 - u_4)^2\left[u_1^2(u_1 + u_3 + u_4 - C)^2 - \frac{4B}{u_3 u_4}u_1\right].$$

The last expression is an algebraic equation of the 4th degree with respect to $u_1$; by solving it and thus expressing the function $u_1$ through the functions $u_3$ and $u_4$, we can reduce system (7) to a system of two differential equations with respect to the desired functions $u_3$ and $u_4$.

Let us assume that at the initial moment of time $t_0 \in \mathbb{R}$ *one of the sought functions is equal to zero.* Due to the symmetry of the system, for convenience we choose the condition $u_2(t_0) = 0$. Then, due to the certain linearity of the differential equations in system (7), the identity $u_2 \equiv 0$ is satisfied, and system (7) takes the form

$$\begin{cases} u_1' = -u_1 u_5 \\ u_3' = u_3 u_4 \\ u_4' = u_4(u_5 - u_3) \\ u_5' = u_5(u_1 - u_4) \end{cases}. \qquad (13)$$

Therefore, from (9), (10) and the condition $u_2 \equiv 0$ we obtain the identities

$$u_4 = \frac{u_3^2 - C u_3 + D}{u_1 - u_3}, \quad u_5 = \frac{C u_1 - u_1^2 - D}{u_1 - u_3}. \qquad (14)$$

From which it follows that system (13) can be reduced to a system of two differential equations with respect to the sought functions $u_1$ and $u_3$. Furthermore, system (13) can be reduced to a single second-order differential equation. To do this, we substitute the expression for $u_5$ from (14) into the first equation of system (13), resulting in the equality

$$u_3 = \frac{u_1(Cu_1 - u_1^2 - D)}{u_1'} + u_1. \tag{15}$$

Substituting the expressions for $u_4$ from (14) and for $u_3$ from (15) into the second equation of system (13), we obtain the identity

$$u_1'' = (3u_1 - C)u_1' - u_1(u_1^2 - Cu_1 + D). \tag{16}$$

The last equation has constant solutions $u_1 = h$, where the constant $h$ is found from the equality $h(h^2 - Ch + D) = 0$. Using the change of variables $u_1 = y$ for convenience, we write equation (16) in the form

$$y'' = (3y - C)y' - y(y^2 - Cy + D).$$

Let's make the usual change of variables $y' = p(y)$ in the last equation:

$$p' = -y(y^2 - Cy + D)p^{-1} + 3y - C. \tag{17}$$

Equation (17) is a special case of the Chini equation (see [15], p. 300, paragraph 1.55)

$$\frac{dy}{dx} = f(x)y^n + g(x)y + h(x), \ n \in \mathbb{Z},$$

for which it is known: *if the quantity*

$$\alpha = f(x)^{-n-1}h(x)^{-2n+1}\left((nf(x)g(x) + f'(x))h(x) - f(x)h'(x)\right)^n \tag{18}$$

*is independent of $x$, then the change of variables*

$$y(x) = \left(\frac{h(x)}{f(x)}\right)^{\frac{1}{n}} v(x)$$

*leads to an equation with separable variables.* In equation (17),

$$f(y) = -y(y^2 - Cy + D), \ g(y) = 0, \ h(y) = 3y - C, \ n = -1;$$

therefore, substituting these functions $f$, $g$, $h$ and the parameter $n$ in (18), for equation (17), we obtain the equality

$$\alpha = -\frac{9}{2}\left(y - \frac{C}{3}\right)^3\left[\left(y - \frac{C}{3}\right)^3 + \frac{C}{6}\left(\frac{2C^2}{9} - D\right)\right]^{-1}.$$

If in the last expression

$$D = \frac{2C^2}{9}, \tag{19}$$

then $\alpha = -\frac{9}{2}$. For specific values of the parameters $C$ and $D$ that satisfy condition (19), equation (17) can be solved using the WolframAlpha Pro computer program.

**5. Conclusion.** In some cases, first integrals of a Volterra chain allow one to find differential equations for the unknown functions [7, 11], significantly simplifying the study of system (1). In the third part of this paper, a new approach to calculating first integrals of a Volterra chain is proposed, which is also applicable to other integrable discrete systems for which there exist changes of variables that allow one to obtain solutions of the Toda chain from solutions of these systems.

In the fourth part of this article, using first integrals to study a Volterra chain with period five, it is shown that the dimension of system (1) can be reduced to two unknown functions; calculating and solving differential equations for the unknown functions required imposing additional conditions on the system.

This work was supported by a grant from the Simons Foundation (1030291, B.D.V.).